\documentclass[12pt]{article}
\usepackage{amssymb}
\begin{document}

\begin{center}
{\Large {\bf Nonrelativistic theory of electroscalar field and Maxwell electrodynamics}}

\vspace{1cm}

{\it D.V. Podgainy$^1$, O.A. Zaimidoroga$^2$}

\vspace{0.5cm}

Joint Institute for Nuclear Research,\\
141980, Dubna, Russia

$^1$ podgainy@jinr.ru\\

$^2$ zaimidor@jinr.ru \\

\end{center}

\vspace{1cm}

{\small In this work, a non-relativistic theory of the electroscalar field being an expansion of the classical Maxwell's electrodynamics is presented. Expansion of the classical electrodynamics is based on the hypothesis about an existing new 4-scalar potential complementary to the 4-vector electrodynamic potential. 4-scalar potential, from the viewpoint of the quantum field concept, describes massless scalar particles with a zero spin whose superposition realizes the Coulomb field. In a nonrelativistic approximation this hypothesis leads to the theory in which along with the electric and magnetic fields there arises a new scalar field that can propagate jointly with the electric field in vacuum in the form of a longitudinal electroscalar wave and, thus, transport of the Coulomb field is carried out. Indicative of the arising complementary scalar field is the analogy between the linear theory of elasticity and Maxwell electrodynamics considered in this work. From this analogy it follows that full agreement between the Lame equations and Maxwell equations is reached under the condition of incompressibility of the elastic continuum, while describing the compressible continuum necessitates introduction of a new scalar field. Since the 4-vector electrodynamic and new 4-scalar potentials do not form a single geometric object in the Minkowski space-time, in a nonrelativistic approximation the electromagnetic and electroscalar fields appear to be independent and do not interfere. In the paper, the problem of interaction of the introduced scalar field with charges and currents is also considered and electrodynamics based on the Fock and Podolsky Lagrangian is briefly discussed.}

\vspace{1cm}

\vspace{0.5cm}

\section*{INTRODUCTION}

As is known, one of the key peculiarities of classic Maxwell equations is the lack of solutions to these equations corresponding to longitudinal waves in vacuum, i.e. waves with the electric-field vector or magnetic-field vector collinear to the wave vector (or to the Pointing vector, which is equivalent). However, the last years have seen a growing interest to theoretic research into the possibility of introducing longitudinal electroscalar waves \cite{waser, Vlaenderen}, and magnitoscalar waves \cite{Khvorostenko} as well as efforts to observe them experimentally \cite{Monstein, Ignatiev}. One of the motivations for expanding classical Maxwell electrodynamics by way of introducing an additional longitudinal mode into this theory is the method, known from quantum electrodynamics, of describing the Coulomb interaction of charged particles by means of longitudinal photons \cite{Dirac1}, although the latter ought to be considered "non-physical" (virtual) because, otherwise, quantum theory will face a number of unsurmountable difficulties \cite{Dirac2}. Thus, despite the "non-physical" character of the longitudinal waves in quantum electrodynamics, one cannot do without introducing such an object; in the meantime, in classical electrodynamics the notion of longitudinal waves is absent in principle and the Coulomb interaction is transmitted instantly. 
On the other side, well-known is the analogy between Maxwell equations and linear elasticity equations. Perfect analogy in this regard is reached under the assumption that elastic continuum is incompressible (complete analogy is reached between theory of elasticity and electrodynamics based on lagrangian of Fock-Podolsky \cite{Fok}, see Appendix). In such a case, elastic continuum or electromagnetic vacuum can be understood as a representation, elaborated in the framework of quantum electrodynamics, about physical vacuum as a plasma consisting of virtual electrons and positrons \cite{Dirac2}. As is known, both transverse and longitudinal or Langmuir waves can propagate in such a plasma. Supposing, while employing a mechanical analogy, that the potentials of the electromagnetic field play the role of displacements in such a medium, one has to admit that the macroscopic Maxwell theory is not in full agreement with the representations following from the quantum theory, or, in other words, Maxwell's theory is an approximate theory in terms of the mechanical analogy and takes no account of compressibility of the electromagnetic vacuum.

    In this paper, a supposition is made that the Coulomb field in terms of the quantum field concept is a superposition of scalar photons, i.e. a superposition of massless scalar particles with a zero spin rather than a superposition of vector photons with a zero spin projection (a detailed consideration of the relation between the longitudinal waves and Coulomb interaction in quantum electrodynamics can be found, e.g., in \cite{Dirac2}). Meanwhile, it is assumed that the introduced scalar particles realize the physical states of the field, i.e. they are observables.  A physical consequence of this hypothesis is arising in the macroscopic electrodynamics of a new longitudinal wave responsible for the transport of the Coulomb field in vacuum. It is evident that such a wave must be observable and, like the transverse electromagnetic wave, must transfer energy and momentum. Since the stated hypothesis requires changing of the classical Maxwell equations, this work will focus on establishing non-contradictoriness of the proposed theoretical scheme precisely in the area of non-relativistic physics.

\section{An analogy with the linear elasticity theory}

Building of theoretical scheme will be started with consideration of an analogy between the linear elasticity theory equations and classic Maxwell equations. Let us note that it is the elastic continuum that supports propagation of both longitudinal and transverse waves, therefore, this analogy will give us a constructive indication of how the equations of generalized electrodynamics should look like.

The basic equation of the linear elasticity theory is the Lame equation, which takes the following form in the absence of outside forces \cite{Lyav}:
\begin{eqnarray}\nonumber
-\ddot{\textbf{u}}+c_{l}^2\nabla{\rm div}{\textbf{u}}-c_{t}^2{\textbf{rot}}{\textbf{rot}}{\textbf{u}}=0,
\end{eqnarray}
where the vector $\textbf{u}$ represents the vector of displacements in an elastic medium, $c_{l}$  and $c_{t}$ are the velocities of propagation of the longitudinal and transverse waves, correspondingly. The displacement vector is the main variable in the linear elasticity theory, although it is not a directly observable quantity. Physically observable quantities in this theory are first-order derivatives of $\textbf{u}$, i.e. $\dot\textbf{u}$ is the velocity of elastic displacements and $\sigma_{ij}=\lambda{\rm div}{\textbf{u}}+2\mu\,u_{ij}$ is the stress tensor, where $\lambda$ and $\mu$ are the elastic constants of the medium and $u_{ij}=1/2(\partial u_i/\partial x_j+\partial u_j/\partial x_i)$ is the elastic deformation tensor \cite{LLTE}. With these variables the Lame equation looks as follows:
\begin{eqnarray}\nonumber
-\frac{\partial^2u_i}{\partial t^2}+\frac{\partial \sigma_{ik}}{\partial x_k}=0.
\end{eqnarray}
However, the Lame equation for the observables presented in this form disguises the structure of wave processes that might arise in an elastic medium, namely, from this form one cannot see arrangement of the wave processes into transverse and longitudinal elastic waves. In order to separate the wave processes explicitly, the following designations are introduced \cite{Dubrovskii}:
\begin{eqnarray}
\textbf{E}=-\dot\textbf{u},\quad\quad\textbf{H}=c_t{\textbf{rot}}{\textbf{u}},\quad\quad W=c_l{\rm div}{\textbf{u}}, 
\label{eq1}
\end{eqnarray}
In this case, the Lame equation takes the form:
\begin{eqnarray}\nonumber
\dot\textbf{E}+c_l\nabla W-c_t{\textbf{rot}}{\textbf{H}}=0.
\end{eqnarray}
Now, let us employ the rotor and divergence operation to the vector \textbf{E}:
\begin{eqnarray}\nonumber
&&\textbf{rot}\textbf{E}=-\textbf{rot}\dot\textbf{u}=-\frac{1}{c_t}\dot\textbf{H},\\[0.2cm]\nonumber
&&{\rm div}\textbf{E}=-{\rm div}\dot\textbf{u}=-\frac{1}{c_l}{\dot W}.
\end{eqnarray}
Thus, for the introduced field denotations one can obtain from the Lame equation the following system of equations:
\begin{eqnarray}\nonumber
&&\dot\textbf{E}+c_l\nabla W-c_t{\textbf{rot}}{\textbf{H}}=0,\\[0.2cm]\label{eq2}
&&\frac{1}{c_t}\dot\textbf{H}+\textbf{rot}\textbf{E}=0,\\[0.2cm]\nonumber
&&\frac{1}{c_l}{\dot W}+{\rm div}\textbf{E}=0,\\[0.2cm]\nonumber
&&{\rm div}\textbf{H}=0,
\end{eqnarray}
which, in the event of $W=c_l{\rm div}\textbf{u}=0$, corresponding to the incompressible elastic continuum, coincides with the system of Maxwell equations in vacuum
\begin{eqnarray}\label{eq3}
&&\dot\textbf{E}-c_t\textbf{rot H}=0,\\[0.2cm]\nonumber
&&\frac{1}{c_t}{\dot \textbf{H}}+\textbf{rot E}=0.
\end{eqnarray}
If the continuous medium does not support rotating motion, for which $\textbf{rot}\textbf{u}=0$ (e.g. liquid or gas), then the system (\ref{eq2}) takes this form:
\begin{eqnarray}\label{eq4}
&&\dot\textbf{E}+c_l\nabla W=0,\\[0.2cm]\nonumber
&&\frac{1}{c_l}{\dot W}+{\rm div}\textbf{E}=0.
\end{eqnarray}
Such a system describes propagation of longitudinal (acoustic) waves in a continuous medium. Thus, the wave degrees of freedom of the elastic continuum are described by two systems of equations via the introduced field definitions (\ref{eq1}), with the vector fields from (\ref{eq3}) being vortical, i.e. ${\rm div}\textbf{E}={\rm div}\textbf{H}=0$, and the field $\textbf{E}$ from (\ref{eq4}) being potential, i.e. $\textbf{rot~E}=0$. The total vector field $\textbf{E}$ will be defined as a sum of the vortical field $\textbf{E}_\bot$, which presents the solution to (\ref{eq3}), and potential field $\textbf{E}_{||}$ determined from (\ref{eq4}). Because the transverse and longitudinal waves propagate in the elastic continuum with different velocities, $c_{t}$ for the transverse and $c_{l}$ for the longitudinal ones, the fields $\textbf{E}_\bot$ and $\textbf{E}_{||}$ can be considered as independent. 

If this analogy is applied to Maxwell's electrodynamics, in other words, if electrodynamic potentials are considered as values characterizing displacements in elastic electromagnetic vacuum and electromagnetic fields are understood as tensions in vacuum corresponding to these displacements, one can expect appearance two field pairs. One of them will be describe strictly transverse wave which correspond electromagnetic wave of classical electrodynamics. While other pair will be describe propagation of longitudinal wave. This field pair will be consist from potential part of electric field (because always ${\rm div}\textbf{H}=0$, that means absence of magnetic charge) and from a new three dimential scalar field, which one introduce in next section.

\section{Scalar field}

As has been noted above, the Coulomb field is represented in quantum electrodynamics as a superposition of photons with a spin projection equal to zero. Such photons are considered non-physical \cite{Fermi32} in the framework of QED. However, one can accept an alternative possibility, i.e. by assuming that the Coulomb field is a superposition of massless scalar particles. Constructively, this supposition is expressed by that along with the 4-vector potential $A^\mu$, submitting in vacuum to the equations 
\begin{eqnarray}
\square A^\mu=0,\quad\quad\partial_\mu A^\mu=0,
\label{eq5}
\end{eqnarray}
where $\square= -\partial^2/c^2\partial t^2+\Delta$ is the d'Alembert operator, we add the scalar field $\lambda$, which is the solution to the vacuum equation: 
\begin{eqnarray}
\square \lambda=0.
\label{eq6}
\end{eqnarray}
Let us introduce the following field definitions: 
\begin{eqnarray}\label{eq7}
&&\textbf{E}_\bot=-\nabla\varphi-\frac{1}{c}\frac{\partial\textbf{A}}{\partial t},\quad\quad
\textbf{H}=\textbf{rot A}\\[0.2cm]\nonumber
&&
\textbf{E}_{||}=\nabla\lambda,\quad\quad W=-\frac{1}{c}\frac{\partial\lambda}{\partial t},
\end{eqnarray}
where $\varphi$ and $\textbf{A}$ are the time and space parts of the electromagnetic 4-potential. It should be noted here that by virtue of their definition the values $\textbf{E}_{||}$ and $W$ are the components of a 4-vector in the Minkowski space-time. Definitions (\ref{eq7}) allow one to obtain from equations (\ref{eq5}) and (\ref{eq6}) two systems of equations, one for the fields $\textbf{E}_\bot$ and $\textbf{H}$
\begin{eqnarray}\nonumber
&&-\frac{1}{c}\frac{\partial\textbf{E}_\bot}{\partial t}+\textbf{rot H}=0,\\[0.2cm]\label{eq8}
&&\frac{1}{c}\frac{\partial\textbf{H}}{\partial t}+\textbf{rot E}_\bot=0,\\[0.2cm]\nonumber
&&{\rm div}\textbf{H}=0,
\end{eqnarray}
and another for the fields $\textbf{E}_{||}$ and $W$:
\begin{eqnarray}\nonumber
&&\frac{1}{c}\frac{\partial\textbf{E}_{||}}{\partial t}+\nabla\,W=0,\\[0.2cm]\label{eq9}
&&\frac{1}{c}\frac{\partial W}{\partial t}+{\rm div}\textbf{E}_{||}=0,\\[0.2cm]\nonumber
&&\textbf{rot E}_{||}=0.
\end{eqnarray}

When electric charge and current are absent, the time component $\varphi$ of the 4-potential can be eliminated with the help of gauge transformations \cite{LLTF}. In this case, the field $\textbf{E}_\bot$ will be defined only via the space component of the four-vector potential $\textbf{A}$: $\textbf{E}_\bot=-\partial\textbf{A}/\partial (c\,t)$ which, in its turn, will satisfy the Coulomb condition: ${\rm div}\textbf{A}=0$. In that event, system (\ref{eq8}) describes propagation of strictly transverse waves (a proof of that fact can be found for example in \cite{LLTF}). System (\ref{eq9}), on the contrary, describes propagation of longitudinal waves.
Indeed, let us consider solution of this system in the form of plane waves $\textbf{E}_{||}=\textbf{E}_0\exp[i(\omega t+\textbf{kr})], W=W_0\exp[i(\omega t+\textbf{kr})]$, where $\textbf{E}_0$ and $W_0$ are the amplitudes, $\textbf{k}$ is the wave vector defining the direction of propagation of a wave with the frequency $\omega$. Then, from (\ref{eq9}) the following is obtained:
\begin{eqnarray}\nonumber
\frac{\omega}{c}\textbf{E}_0+\textbf{k}W_0=0,\quad\quad \frac{\omega}{c}W_0+(\textbf{kE}_0)=0,\quad\quad \textbf{k}\times\textbf{E}_0=0, 
\end{eqnarray}
i.e. the vector of the electric field vibrates along the direction of wave propagation. The dispersion relation for the given wave has the form $\omega=kc$.  

For obtaining the energy conservation law for fields $\textbf{E}_{||}$ and $W$, let us multiply first equation (\ref{eq9}) scalarwise by $\textbf{E}_{||}$ and the second equation by $W$
\begin{eqnarray}\nonumber
&&\frac{1}{c}\textbf{E}_{||}\frac{\partial\textbf{E}_{||}}{\partial t}+\textbf{E}_{||}\nabla\,W=0,\\[0.2cm]\nonumber
&&\frac{1}{c}W\frac{\partial W}{\partial t}+W{\rm div}\textbf{E}_{||}=0.
\end{eqnarray}
Placing $\textbf{E}_{||}$ and $W$ under the time derivative sign, summing up the obtained equations and integrating over volume, one obtains:
\begin{eqnarray}\nonumber
\frac{1}{8\pi}\frac{\partial}{\partial t}\int(\textbf{E}_{||}^2+W^2)d\,V+
\frac{c}{4\pi}\int(\textbf{E}_{||}\nabla W+W{\rm div}\textbf{E}_{||})d\,V=0.
\end{eqnarray}
The expression $\textbf{E}_{||}\nabla W+W{\rm div\textbf{E}_{||}}$ equals ${\rm div}(W\textbf{E}_{||})$ and the second integral on the left can be reduced to the surface integral. Introducing denotations for the energy density 
\begin{eqnarray}
\epsilon_{EW}=\frac{\textbf{E}_{||}^2+W^2}{8\pi},
\label{eq10}
\end{eqnarray}
and for the flow vector 
\begin{eqnarray}
\textbf{s}_{EW}=\frac{c}{4\pi}\textbf{E}_{||}W,
\label{eq11}
\end{eqnarray}
one obtains:
\begin{eqnarray}
\frac{\partial}{\partial t}\int\epsilon_{EW}d\,V+\oint\textbf{s}_{EW}\textbf{d}\sigma=0,
\end{eqnarray}
where $\textbf{d}\sigma$ is an element of the area of surface of volume $V$. From the definition of the vector $\textbf{s}_{EW}$ it follows that the energy flow in the plane wave considered above is collinear to the vector of the field $\textbf{E}_{||}$, which is, in its turn, collinear to the wave propagation vector $\textbf{k}$, in other words, system (\ref{eq9}) describes propagation of longitudinal waves in vacuum.

As is seen, systems of equations (\ref{eq8}) and (\ref{eq9}) and systems (\ref{eq3}) and (\ref{eq4}), obtained to describe propagation of elastic waves, are truly identical and coincide, if we set $c_t=c_l=c$ (as we note above, complete analogy is reached between theory of elasticity and electrodynamics based on lagrangian of Fock-Podolsky, see Appendix). 
Let us note here that equations (\ref{eq3}) and (\ref{eq4}) are obtained from the Lame equation written for one three-dimensional vector potential $\textbf{u}$  and, thus, the fields defined in (\ref{eq1}) are coupled. In our case, because the potentials $A^\mu$ and $\lambda$ are not related by any correlation (potentials $A^\mu$ and $\lambda$ transform independently when Lorentz transformations, i.e. $A^\mu$ and $\lambda$ non form geometrical object in Minkowski space-time), the couples of fields $(\textbf{E}_\bot, \textbf{H})$ and $(\textbf{E}_{||}, W)$ are independent, and we cannot say that fields $\textbf{E}_\bot$ and $\textbf{E}_{||}$ are vortex and potential parts of electric field $\textbf{E}$. For discover correlation between $\textbf{E}_\bot$ and $\textbf{E}_{||}$ is necessary to investigate behavior of these fields in presence of charges and currents. 

\section{Scalar field in presence of charges and currents}

\subsection{Fields equations in presence of charges and currents}

By consideration of a question on interaction of fields with charges and currents, we suppose that the time component of the 4-potential $A^0=\varphi$ equals zero, and 3-vector $\textbf{A}$ satisfies the Coulomb condition ${\rm div}\textbf{A}=0$. Let us
write down the vacuum equations for potentials:
\begin{eqnarray}\label{eq13}
&&-\frac{1}{c^2}\frac{\partial^2\textbf{A}}{\partial t^2}+\Delta\textbf{A}=0,\\[0.2cm]\nonumber
&&-\frac{1}{c^2}\frac{\partial^2\lambda}{\partial t^2}+\Delta\lambda=0.
\end{eqnarray}
In order to derive potential equations for the case with presence of charges, let us resort to the Lagrangian formalism. Three-dimensional Lagrangians corresponding to (\ref{eq13}) have the form:
\begin{eqnarray}\nonumber
&&L_{EH}=\frac{\textbf{E}_\bot^2-\textbf{H}^2}{8\pi}=\frac{1}{8\pi}\left(\frac{1}{c}\frac{\partial\textbf{A}}{\partial t}\right)^2-\frac{1}{8\pi}(\textbf{rot A})^2,\\[0.2cm]\nonumber
&&L_{EW}=\frac{W^2-\textbf{E}_{||}^2}{8\pi}=\frac{1}{8\pi}\left(\frac{1}{c}\frac{\partial\lambda}{\partial t}\right)^2-\frac{1}{8\pi}(\nabla\lambda)^2.
\end{eqnarray}
Let us introduce the following interaction $(\textbf{Aj})/c$ for $L_{EH}$ and $\rho\lambda$ for $L_{EW}$:
\begin{eqnarray}\nonumber
&&L_{EH}=\frac{\textbf{E}_\bot^2-\textbf{H}^2}{8\pi}=\frac{1}{8\pi}\left(\frac{1}{c}\frac{\partial\textbf{A}}{\partial t}\right)^2-\frac{1}{8\pi}(\textbf{rot A})^2+\frac{1}{c}\textbf{Aj},\\[0.2cm]\nonumber
&&L_{EW}=\frac{W^2-\textbf{E}_{||}^2}{8\pi}=\frac{1}{8\pi}\left(\frac{1}{c}\frac{\partial\lambda}{\partial t}\right)^2-\frac{1}{8\pi}(\nabla\lambda)^2+\rho\lambda,
\end{eqnarray}
then system (\ref{eq13}) takes the form:
\begin{eqnarray}\label{eq14}
&&-\frac{1}{c^2}\frac{\partial^2\textbf{A}}{\partial t^2}+\Delta\textbf{A}=-\frac{4\pi}{c}\textbf{j},\\[0.2cm]\nonumber
&&-\frac{1}{c^2}\frac{\partial^2\lambda}{\partial t^2}+\Delta\lambda=-4\pi\rho,
\end{eqnarray}
where $\rho$ and $\textbf{j}$ are the charge and current densities. Since the vector potential $\textbf{A}$ satisfies the Coulomb condition, (\ref{eq14}) includes only the vortex part of the transport current, i.e. ${\rm div}\textbf{j}=0$. Thus,  the continuity equation 
\begin{eqnarray}
\frac{\partial\rho}{\partial t}+{\rm div}\textbf{j}=0, 
\label{eq15}
\end{eqnarray}
no impose constraints on potentials $\textbf{A}$ and $\lambda$.

Substitution in (\ref{eq14}) of definitions of fields (\ref{eq7}) gives:
\begin{eqnarray}\nonumber
&&-\frac{1}{c}\frac{\partial\textbf{E}_\bot}{\partial t}+\textbf{rot H}=\frac{4\pi}{c}\textbf{j},\\[0.2cm]\label{eq16}
&&\frac{1}{c}\frac{\partial\textbf{H}}{\partial t}+\textbf{rot E}_\bot=0,\\[0.2cm]\nonumber
&&{\rm div}\textbf{H}=0,
\end{eqnarray}
and 
\begin{eqnarray}\nonumber
&&\frac{1}{c}\frac{\partial\textbf{E}_{||}}{\partial t}+\nabla\,W=0,\\[0.2cm]\label{eq17}
&&\frac{1}{c}\frac{\partial W}{\partial t}+{\rm div}\textbf{E}_{||}=-4\pi\rho,\\[0.2cm]\nonumber
&&\textbf{rot E}_{||}=0.
\end{eqnarray}
It is seen from these systems that the electrostatics equation, ${\rm div}\textbf{E}_{||}=-4\pi\rho$, follows from system (\ref{eq17}) and magnetostatics one, $\textbf{rot H}=4\pi\textbf{j}/c$, follows from (\ref{eq16}). These two systems of equations (\ref{eq16}) and (\ref{eq17}) can be rewritten in the form of wave equations for the fields:
\begin{eqnarray}\label{eq18}
&&-\frac{1}{c^2}\frac{\partial^2\textbf{E}_\bot}{\partial t^2}+\Delta\textbf{E}_\bot=\frac{4\pi}{c^2}\frac{\partial\textbf{j}}{\partial t},\quad\quad {\rm div}\textbf{E}_\bot=0, \\[0.2cm]\nonumber
&&-\frac{1}{c^2}\frac{\partial^2\textbf{H}}{\partial t^2}+\Delta\textbf{H}=-\frac{4\pi}{c^2}\textbf{rot j},\quad\quad {\rm div}\textbf{H}_\bot=0;
\end{eqnarray}
and 
\begin{eqnarray}\label{eq19}
&&-\frac{1}{c^2}\frac{\partial^2\textbf{E}_{||}}{\partial t^2}+\Delta\textbf{E}_{||}=-4\pi\nabla\rho,\quad\quad \textbf{rot E}_{||}=0, \\[0.2cm]\nonumber
&&-\frac{1}{c^2}\frac{\partial^2 W}{\partial t^2}+\Delta W=\frac{4\pi}{c}\frac{\partial\rho}{\partial t}.
\end{eqnarray}
The identity $\Delta\equiv-\textbf{rot rot}+\nabla{\rm div}$ being used while deriving these equations. It is seen from these equations that the time-varying current with a rotor different from zero is the source of transverse waves, while the time-dependent non-uniform charge density serves as the source of longitudinal waves. Here, taking into account the continuity equation (\ref{eq15}) the latter equation (\ref{eq19}) can be presented in the form: 
\begin{eqnarray}\nonumber
&&-\frac{1}{c^2}\frac{\partial^2 W}{\partial t^2}+\Delta W=-\frac{4\pi}{c}{\rm div}\textbf{j}.
\end{eqnarray}
It is seen from (\ref{eq17}) and (\ref{eq19}) that longitudinal electroscalar wave respond to transport of the Coulomb field.

The energy conservation laws for (\ref{eq16}) and (\ref{eq17}) have the form:
\begin{eqnarray}\label{eq20}
&&\frac{\partial}{\partial t}\int\epsilon_{EH}d\,V+\oint\textbf{s}_{EH}\textbf{d}\sigma=-\int(\textbf{jE}_\bot)d\,V,\\[0.2cm]
&&\frac{\partial}{\partial t}\int\epsilon_{EW}d\,V+\oint\textbf{s}_{EW}\textbf{d}\sigma=-c\int(\rho W)d\,V,
\label{eq21}
\end{eqnarray}
here $\epsilon_{EW}$ and $\textbf{s}_{EW}$ in (\ref{eq21} defined by (\ref{eq10}) and (\ref{eq11}) correspondingly, and electromagnetic energy density $\epsilon_{EH}$ and flux vector $\textbf{s}_{EH}$ have the form:
\begin{eqnarray}\nonumber
\epsilon_{EH}=\frac{\textbf{E}_\bot^2+\textbf{H}}{8\pi},\quad\quad \textbf{s}_{EH}=\frac{c}{4\pi}\left[\textbf{E}_\bot\times\textbf{H}\right].
\end{eqnarray}
The quantities on the right sides of (\ref{eq20}) and (\ref{eq21}) determine the energy dissipation of the fields per unit time while they are interacting with charges and currents. Of particular interest is the integrand $\rho\,W$ from the right side of (\ref{eq21}). Using the definition $W=-\partial\lambda/\partial (c\,t)$ and continuity equation (\ref{eq15}) this expression can be brought to the following form:
\begin{eqnarray}
\rho\,W=-\rho\frac{1}{c}\frac{\partial\lambda}{\partial t}=-\frac{\partial}{\partial t}\left(\frac{1}{c}\lambda\rho\right)+\frac{1}{c}\lambda\frac{\partial\rho}{\partial t}=-\frac{\partial}{\partial t}\left(\frac{1}{c}\lambda\rho\right)-{\rm div}\left(\frac{1}{c}\lambda\,\textbf{j}\right)+\frac{1}{c}\textbf{j}\nabla\lambda.
\end{eqnarray}
Owing to the definition $\textbf{E}_{||}=\nabla\lambda$, (\ref{eq21}) can be represented in the form:
\begin{eqnarray}
\frac{\partial}{\partial t}\int(\epsilon_{EW}-\rho\lambda)d\,V+\oint(\textbf{s}_{EW}-\lambda\textbf{j})\textbf{d}\sigma=-\int(\textbf{j}\textbf{E}_{||})d\,V.
\end{eqnarray}
The first integral presents the total energy of the system of fields $\textbf{E}_{||}$ and $W$ interacting with charges. 
The summand $\lambda\textbf{j}$ has been added to the surface integral. That summand can be interpreted as energy transfer at the expense of motion of charges, indeed, taking into account that $\textbf{j}=\rho\,\textbf{v}$, one obtains $\lambda\textbf{j}=(\rho\lambda)\textbf{v}.$ In other words, the total flux consists of two components - the first one is responsible for the energy transfer via electroscalar radiation and the other is connected with the mechanical emission of charges from the area limited by the integration surface.
	
\subsection{Charged particle in electromagnetic and electroscalar fields}

	Let us now consider the behavior of a charged particle in presence of the introduced fields. In the classical theory, the Lagrangian of the motion of a charged body has the form:
\begin{eqnarray}
L=\frac{m\textbf{v}^2}{2}+\frac{1}{c}(q\textbf{A}\textbf{v})-q\lambda,
\label{eq23}
\end{eqnarray}
where $q$, $m$ and $\textbf{v}$ are the body's charge, mass and velocity correspondingly and instead of potential $\varphi$ one introduced potential $\lambda$. Since electromagnetic and electroscalar fields are independent in the developed theoretical scheme, we can consider their action on charges also individually. 

{\sf Equations of motion of charged particle in transverse fields.} Let us consider the electromagnetic interaction in the first place. In this case, Lagrangian (\ref{eq23}) retains a component containing the vector potential $\textbf{A}$:
\begin{eqnarray} 
L=\frac{m\textbf{v}^2}{2}+\frac{1}{c}(q\textbf{A}\textbf{v}).
\label{eq24}
\end{eqnarray}
An equation of motion for a charged body will be now obtained by substituting (\ref{eq24}) in the Euler-Lagrange formula:
\begin{eqnarray} 
\frac{d}{dt}\frac{\partial\,L}{\partial\textbf{v}}=\frac{\partial\,L}{\partial\textbf{r}}.
\label{eq25}
\end{eqnarray}
The derivative $\partial\,L/\partial\textbf{v}$ defines the generalized momentum $\textbf{P}$ of a charge in the electromagnetic field:
\begin{eqnarray} 
\textbf{P}=\frac{\partial\,L}{\partial\textbf{v}}=m\textbf{v}+\frac{q}{c}\textbf{A}.
\end{eqnarray}
Calculation of this expression of a total time derivative gives the following:
\begin{eqnarray}
\frac{d}{dt}\frac{\partial\,L}{\partial\textbf{v}}=m\frac{d\textbf{v}}{dt}+\frac{q}{c}\frac{d\textbf{A}}{dt}=m\frac{d\textbf{v}}{dt}+\frac{q}{c}\frac{\partial\textbf{A}}{\partial\,t}+\frac{q}{c}(\textbf{v}\nabla)\textbf{A}=m\frac{d\textbf{v}}{dt}+\frac{q}{c}(\textbf{v}\nabla)\textbf{A}-q\textbf{E}_\bot.
\label{eq26}
\end{eqnarray}
Here we have revealed the total time derivative of the vector $\textbf{A}$ using the formula:
\begin{eqnarray}\nonumber
\frac{d}{dt}=\frac{\partial}{\partial\,t}+(\textbf{v}\nabla).
\end{eqnarray}
Computation of the right side of (\ref{eq25}) for Lagrangian (\ref{eq24}) gives the following:
\begin{eqnarray}
\frac{\partial\,L}{\partial\textbf{r}}=\frac{q}{c}\nabla(\textbf{v}\textbf{A})=\frac{q}{c}(\textbf{v}\nabla)\textbf{A}+\frac{q}{c}\left[\textbf{v}\times\textbf{rot A}\right]=\frac{q}{c}(\textbf{v}\nabla)\textbf{A}+\frac{q}{c}\left[\textbf{v}\times\textbf{H}\right].
\label{eq27}
\end{eqnarray}
By substituting (\ref{eq26}) and (\ref{eq27}) in (\ref{eq25}), the final expression for the equation of motion in transverse fields is obtained:
\begin{eqnarray}
m\frac{d\textbf{v}}{dt}=q\textbf{E}_\bot+\frac{q}{c}\left[\textbf{v}\times\textbf{H}\right].
\label{eq28}
\end{eqnarray}
The obtained expression fully coincides with the classical one despite the fact that (\ref{eq23}) was used instead of Lagrangian (\ref{eq24}). Let us now calculate the Hamiltonian $H$: 
\begin{eqnarray}
H=\frac{\partial\,L}{\partial\textbf{v}}\textbf{v}-L.
\label{eq29}
\end{eqnarray}
In the given case this Hamiltonian has the following form:
\begin{eqnarray}
H=\left(m\textbf{v}+\frac{q}{c}\textbf{A}\right)\textbf{v}-\left(\frac{m\textbf{v}^2}{2}+\frac{1}{c}(q\textbf{A}\textbf{v})\right)=\frac{m\textbf{v}^2}{2}.
\label{eq30}
\end{eqnarray}
Thus, one can see that the electromagnetic field changes the generalized momentum of the charge but does not change the total energy. Let us now compute the change in the kinetic energy of the charged body, multiplying for this purpose (\ref{eq28}) scalarwise by $\textbf{v}$: 
\begin{eqnarray}
\frac{d}{dt}\frac{m\textbf{v}^2}{2}=\textbf{I}\textbf{E}_\bot,
\label{eq31}
\end{eqnarray}
where the designation $\textbf{I}$ for the total current $\textbf{I}=q\textbf{v}$ is introduced. In order to derive the full law of conservation of electromagnetic and kinetic energy, it is required to pass in (\ref{eq20}) from the continuous charge density to the point charge and perform volume integration \cite{Dzhekson}. This passage is done with the help of the formula $\rho=q\delta (\textbf{r}-\textbf{r}')$ where $\delta (\textbf{r}-\textbf{r}')$ is the $\delta$-function. Doing integration and taking into account that the fields tend to zero at infinity, the following can be obtained:
\begin{eqnarray}
\frac{d}{dt}\Xi_{EH}=-\textbf{I}\textbf{E}_\bot,
\label{eq32}
\end{eqnarray}
where $\Xi_{EH}$ denotes the total energy of the electromagnetic field. Summing up (\ref{eq31}) and (\ref{eq32}), one can get 
\begin{eqnarray}
\frac{d}{dt}\left(\Xi_{EH}+\frac{m\textbf{v}^2}{2}\right)=0.
\label{eq33}
\end{eqnarray}
From last formula we can see that total energy is positive definite quantity. 
Taking into consideration (\ref{eq30}), this formula can be given the following form:   
\begin{eqnarray}
\frac{d}{dt}\left(\Xi_{EH}+H\right)=0.
\end{eqnarray}

{\sf Equations of motion of charged particle in longitudinal fields.} Let us now address the problem of interaction of charges with the electroscalar field. The Lagrangian in this case has the form:
\begin{eqnarray}
L=\frac{m\textbf{v}^2}{2}-q\lambda,
\label{eq35}
\end{eqnarray}
and for the Hamiltonian the following expression can be obtained:
\begin{eqnarray}
H=\frac{\partial\,L}{\partial\textbf{v}}\textbf{v}-L=\frac{m\textbf{v}^2}{2}+q\lambda.
\label{eq36}
\end{eqnarray}
Let us note that in contrast to the electromagnetic field the electroscalar one changes the Hamiltonian of the charged body. Now the generalized momentum can be calculated:
\begin{eqnarray}
\textbf{P}=\frac{\partial\,L}{\partial\textbf{v}}=m\textbf{v}.
\end{eqnarray}
It is seen that the generalized momentum coincides with the common mechanical momentum, that is the electroscalar interaction changes the Hamiltonian but does not change the momentum, in other words, longitudinal and transverse fields make an opposite contribution in mechanical momentum and kinetic energy. Let us now derive the Lorentz force, substituting (\ref{eq35}) in (\ref{eq25}):
\begin{eqnarray}
m\frac{d\textbf{v}}{dt}=-q\nabla\lambda=-q\textbf{E}_{||}.
\end{eqnarray}
It follows from this formula that, in the first place, the field $W$ does not enter into the expression for the force acting on the charge in explicit form and, in the second place, the field $\textbf{E}_{||}$ enters into this formula with a negative sign. The law of time evolution of kinetic energy in this case looks as follows:  
\begin{eqnarray}
\frac{d}{dt}\frac{m\textbf{v}^2}{2}=-\textbf{I}\textbf{E}_{||}.
\label{eq37}
\end{eqnarray}
Using the same method that was employed in the case of electromagnetic fields, formula (\ref{eq21}) can be rewritten for one charged body:
\begin{eqnarray}
\frac{d}{dt}\left(\Xi_{EW}-q\lambda\right)=-\textbf{I}\textbf{E}_{||},
\end{eqnarray}
where $\Xi_{EW}$ denotes the total energy of the electroscalar field.
Let us now subtract the obtained result from (\ref{eq37}):
\begin{eqnarray}
\frac{d}{dt}\left(\frac{m\textbf{v}^2}{2}+q\lambda-\Xi_{EW}\right)=0.
\end{eqnarray}
Taking into consideration the Hamiltonian definition for the charge in an electroscalar field (\ref{eq36}), the latter formula can be given the final form:
\begin{eqnarray}
\frac{d}{dt}\left(H-\Xi_{EW}\right)=0.
\end{eqnarray}
Thus, we obtain that the mechanical and electroscalar energy enters into the total energy with different signs.

In conclusion of this section, let us write out once again the Lagrangian for the charge under the action of both the electromagnetic and electroscalar fields:  
\begin{eqnarray}
L=\frac{m\textbf{v}^2}{2}+\frac{1}{c}(q\textbf{A}\textbf{v})-q\lambda.
\end{eqnarray}
Following from this Lagrangian is the equation of motion with a generalized Lorentz force:
\begin{eqnarray}
m\frac{d\textbf{v}}{dt}=q(\textbf{E}_\bot-\textbf{E}_{||})+\frac{q}{c}\left[\textbf{v}\times\textbf{H}\right],
\label{eq38}
\end{eqnarray}
and resulting from the latter formula is the law of variation of the charge kinetic energy:
\begin{eqnarray}
\frac{d}{dt}\frac{m\textbf{v}^2}{2}=\textbf{I}(\textbf{E}_\bot-\textbf{E}_{||}).
\label{eq39}
\end{eqnarray}
Moreover, it is seen from (\ref{eq38}) and (\ref{eq39}) that fields $\textbf{E}_\bot$ and $\textbf{E}_{||}$ enter into expression for generalized Lorentz force and into expression for variation of kinetic energy as linear combination. It allows to conclude that these fields are vortex and potential parts of total electric field $\textbf{E}$. 

The law of conservation of total energy for a charged body has the form: 
\begin{eqnarray}
\frac{d}{dt}\left(H+\Xi_{EH}-\Xi_{EW}\right)=0,
\label{eq40}
\end{eqnarray}
where $H$ implies the total mechanical energy of the charge in the field $H=m\textbf{v}^2/2+q\lambda$. The interesting peculiarity of formula (\ref{eq40}) is that the electromagnetic energy $\Xi_{EH}$ and electroscalar energy $\Xi_{EW}$ enter into it with different signs. This result can be interpreted as a probability of arising of bound states of interacting charges in case dynamical Coulomb field. 

\section*{Conclusion}

The hypothesis about a complementary 4-scalar potential considered in this work leads to a  well-formed three-dimensional nonrelativistic theory describing propagation of both longitudinal electroscalar and transverse electromagnetic waves in vacuum. The peculiarity of the proposed theoretical scheme is that the transverse and longitudinal wave processes in vacuum are described by independent couples of fields, $(\textbf{E}_\bot, \textbf{H})$ and $(\textbf{E}_{||}, W)$, correspondingly. Such division is due to that the 4-vector electromagnetic and introduced 4-scalar potentials do not form a single geometric object in the Minkowski space-time (i.e. they are transformed independently at Lorentz transformations) on the one hand and are not connected by any differential relation on the other hand. That the couples of fields $(\textbf{E}_\bot, \textbf{H})$ and $(\textbf{E}_{||}, W)$ are whole physical field follow from consideration of interaction of the fields with charges and currents. Since in this case the expression for the force acting on the charge and expression for the law of variation of the kinetic energy of a charged particle include the linear combination  $(\textbf{E}_\bot-\textbf{E}_{||})$ that allows one to interpret the fields $\textbf{E}_\bot$  and  $\textbf{E}_{||}$ as transverse and longitudinal components of the electric field. As has been shown, it is the longitudinal electroscalar wave (the couple of fields $(\textbf{E}_{||}, W)$) that is responsible for the transport of the Coulomb field. The immediate source for the new scalar field $W$ is the time-varying charge density or the component of the transfer current density with a divergence different from zero. 

We will note that the proposed theory is not a unique one that allows describing propagation of a longitudinal wave of electromagnetic nature in vacuum. As an illustration, in the Appendix the electrodynamics on the basis of the Fock-Podolsky Lagrangian is considered, which also describes propagation of the longitudinal and transverse waves in vacuum, where after dividing the vector fields into potential and vortical components the obtained equations have the same form as the equations in the proposed formalism. However, the Fock-Podolsky electrodynamics has a number of obstacles in providing a nonrelativistic description of interaction of charges with fields introduced in the framework of this theory. Particularly, the law of conservation of total energy is not fulfilled, the scalar field introduced in the framework of this theory does not contain any sources in apparent form, and the requirement of gauge invariance of the Fock-Podolsky Lagrangian leads to that the charges and currents depend on the choice of calibration. The proposed theory is free from obstacles arising in the given approach. 

We will note in conclusion that the key motivation for constructing the proposed theory has been the existing obstacles in quantum electrodynamics while trying to describe the Coulomb field  \cite{Dirac2}, since the latter can be described only as a superposition of longitudinal photons realizing the non-physical conditions of the field. Such an approach is not quite reasonable as, for instance, the electrostatic field is a totally physical object. Therefore, in the event of creating quantum electrodynamics considering the Coulomb field as a superposition of the observed quanta, in the classical limit it must contain description of the wave transport of the Coulomb field, which is missing in Maxwell's electrodynamics. Notwithstanding the fact that the theory presented in this work is not directed first-hand at overcoming obstacles in quantum electrodynamics, however, it can serve such a classical limit.

\newpage

\section*{Appendix. Electrodynamics on the basis of the Fock-Podolsky Lagrangian.}
\setcounter{equation}{0}

Of the fullest analogy with the linear elasticity theory is the theory based on the Lagrangian introduced by V.A.Fock and B.Podolsky in 1932 for canonical quantization of Maxwell's electrodynamics \cite{Fok}:  
\begin{eqnarray}
\Lambda=\frac{1}{8\pi}\left(\textbf{E}^2-\textbf{H}^2\right)-\frac{1}{8\pi}\left(\frac{1}{c}\frac{\partial\varphi}{\partial t}+{\rm div}\textbf{A}\right)^2+\frac{1}{c}\textbf{Aj}-\rho\varphi,
\label{ap1}
\end{eqnarray}
For the electromagnetic potentials $\varphi$ and $\textbf{A}$ directly following from this Lagrangian are the wave equations:
\begin{eqnarray}
-\frac{1}{c^2}\frac{\partial^2\varphi}{\partial t^2}+\Delta\varphi=-4\pi\rho,\quad\quad
-\frac{1}{c^2}\frac{\partial^2\textbf{A}}{\partial t^2}+\Delta\textbf{A}=-\frac{4\pi}{c}\textbf{j}.
\label{ap2}
\end{eqnarray}
By using the expressions for  $\textbf{E}$ and $\textbf{H}$ via the potentials    
\begin{eqnarray}
\textbf{E}=-\nabla\varphi-\frac{1}{c}\frac{\partial\textbf{A}}{\partial t},\quad\quad \textbf{H}=\textbf{rot\,A},
\label{ap3}
\end{eqnarray}
and introducing the designation
\begin{eqnarray}
W_{FP}=\frac{1}{c}\frac{\partial\varphi}{\partial t}+{\rm div}\textbf{A},
\label{ap4}
\end{eqnarray}
equations (\ref{ap2}) can be presented in the form:
\begin{eqnarray}\nonumber
&&-\frac{1}{c}\frac{\partial\textbf{E}}{\partial t}-\nabla\,W_{FP}+\textbf{rot\,H}=\frac{4\pi}{c}\textbf{j},\\[0.2cm]\label{ap5}
&&\frac{1}{c}\frac{\partial\textbf{H}}{\partial t}+\textbf{rot\,E}=0,\\[0.2cm]\nonumber
&&\frac{1}{c}\frac{\partial\,W_{FP}}{\partial t}+{\rm div}\textbf{E}=4\pi\rho,\\[0.2cm]\nonumber
&&{\rm div}\textbf{H}=0.
\end{eqnarray}
In the vacuum case, system (\ref{ap5}) is fully equivalent to system (\ref{eq2}) of the first section, if we set in the latter $c_l=c_t=c$. Fulfillment of the Lorentz condition for the electromagnetic potentials  $W_{FP}=\partial\varphi/\partial (c\,t)+{\rm div}\textbf{A}=0$ transfers system (\ref{ap5}) into the system of Maxwell's equations, which, in their turn, in the vacuum case describe propagation of strictly transverse waves. Thus, the Lorentz condition for the potentials is a condition of incompressibility of the electromagnetic vacuum, from the viewpoint of the mechanical analogy, being equivalent to the condition  ${\rm div}\textbf{u}=0$ which, in its turn, prohibits the existence of longitudinal (acoustic) waves in an elastic  medium. From the viewpoint of electrodynamics, the Lorentz condition is closely connected with the electric charge conservation law. In order to observe this, let us differentiate the wave equation for the scalar potential with respect to time and apply the divergence operation to the second equation of system (\ref{ap2}):
\begin{eqnarray}\nonumber
&&-\frac{1}{c^2}\frac{\partial}{\partial\,t}\frac{\partial^2\varphi}{\partial t^2}+\frac{\partial}{\partial\,t}\Delta\varphi=-4\pi\frac{\partial\rho}{\partial\,t},\\[0.2cm]\nonumber
&&-\frac{1}{c^2}{\rm div}\frac{\partial^2\textbf{A}}{\partial t^2}+{\rm div}\nabla{\rm div}\textbf{A}=-\frac{4\pi}{c}{\rm div}\textbf{j}.
\end{eqnarray}
Summing up the obtained equations and taking into account that $\Delta\equiv{\rm div}\nabla$ we get the following equation:
\begin{eqnarray}\nonumber
-\frac{1}{c^2}\frac{\partial}{\partial\,t^2}\left(\frac{1}{c}\frac{\partial\varphi}{\partial t}+{\rm div}\textbf{A}\right)+\Delta\left(\frac{1}{c}\frac{\partial\varphi}{\partial t}+{\rm div}\textbf{A}\right)=-\frac{4\pi}{c}\left(\frac{\partial\rho}{\partial\,t}+{\rm div}\textbf{j}\right),
\end{eqnarray}
from which it is seen that for the fulfillment of the continuity equation, the field $W_{FP}$ must satisfy the uniform wave equation:
\begin{eqnarray}\nonumber
-\frac{1}{c^2}\frac{\partial\,W_{FP}}{\partial\,t^2}+\Delta\,W_{FP}=0,
\end{eqnarray}
and the trivial solution to this equation is $W_{FP}=0$, which corresponds to the Lorentz condition. Let us note here that usually the Lorentz condition is introduced while passing from the potential equations corresponding to Maxwell's field equations to the wave equations for potentials \cite{Dzhekson}. System of equations (\ref{ap5}) describes propagation of both transverse and longitudinal waves. This can be seen if we divide the vectors from (\ref{ap5}) into the longitudinal and transverse, relative to propagation, components of the wave:
\begin{eqnarray}\label{ap6}
&&-\frac{1}{c}\frac{\partial\textbf{E}_\bot}{\partial t}+\textbf{rot\,H}=\frac{4\pi}{c}\textbf{j}_\bot,\\[0.2cm]\nonumber
&&\frac{1}{c}\frac{\partial\textbf{H}}{\partial t}+\textbf{rot\,E}_\bot=0;\\[0.2cm]\label{ap7}
&&\frac{1}{c}\frac{\partial\textbf{E}_{||}}{\partial t}+\nabla\,W_{FP}=-\frac{4\pi}{c}\textbf{j}_{||},\\[0.2cm]\nonumber
&&\frac{1}{c}\frac{\partial\,W_{FP}}{\partial t}+{\rm div}\textbf{E}_{||}=4\pi\rho.
\end{eqnarray}
Here $\textbf{E}_\bot$ - is the transverse part of the electric field, while for this field the following equations are fulfilled: $\textbf{E}_\bot=-\partial\textbf{A}_\bot/\partial(c\,t)\,\, {\rm div}\textbf{E}_\bot=0$, where $\textbf{A}_\bot$ - is the transverse part of the vector potential ${\rm div}\textbf{A}_\bot=0$. $\textbf{E}_{||}$ - defines the longitudinal part of the electric field: $\textbf{E}_{||}=-\nabla\varphi-{\partial\textbf{A}_{||}}/{\partial (c\,t)},\,\, \textbf{rot\,E}_{||}=0$, and, correspondingly $\textbf{A}_{||}$ - is the longitudinal part of the vector potential $\textbf{rot\,A}_{||}=0$. The vectors $\textbf{j}_\bot$ and $\textbf{j}_{||}$ are the transverse and longitudinal components of the current density, satisfying the condition:  ${\rm div}\textbf{j}_\bot=0$ and $\textbf{rot\,j}_{||}=0$. Systems (\ref{ap6}) and (\ref{ap7}) describe propagation of the transverse and longitudinal waves respectively.

Let us now come back to system (\ref{ap5}) and determine for it the energy conservation law. For this purpose we multiply the first equation of this system scalarwise by $\textbf{E}$, second one by $\textbf{H}$ and third one by $W_{FP}$:   
\begin{eqnarray}\nonumber
&&\frac{1}{2c}\frac{\partial\textbf{E}^2}{\partial t}-\textbf{E}\,\textbf{rot\,H}+\textbf{E}\nabla\,W_{FP}=-\frac{4\pi}{c}\textbf{jE},\\[0.2cm]
&&\frac{1}{2c}\frac{\partial\textbf{H}^2}{\partial t}+\textbf{H}\,\textbf{rot\,E}=0,\\[0.2cm]\nonumber
&&\frac{1}{2c}\frac{\partial\,W_{FP}^2}{\partial t}+W_{FP}{\rm div}\textbf{E}=4\pi\rho\,W_{FP}.
\label{ap8}
\end{eqnarray}
Summing up the obtained expressions and integrating over volume, we have:
\begin{eqnarray}
\frac{\partial}{\partial t}\int\epsilon\,d\,V+\oint\textbf{s}\textbf{d}\sigma=\int\left(-(\textbf{jE})+c\,\rho\,W_{FP}\right),
\label{ap9}
\end{eqnarray}
here
\begin{eqnarray}
\epsilon=\frac{\textbf{E}^2+\textbf{H}^2+W_{FP}^2}{8\pi},\quad\quad \textbf{s}=\textbf{s}_\bot+\textbf{s}_{||},\quad \textbf{s}_\bot=\frac{c}{4\pi}\left[\textbf{E}\times\textbf{H}\right],\quad \textbf{s}=\frac{c}{4\pi}\textbf{E}\,W_{FP}.
\end{eqnarray}
$\epsilon$ is the density of the total energy, $\textbf{s}$ is the generalized vector of the energy flux density, while $\textbf{s}_\bot$ is the density of the electromagnetic energy flux density in transverse wave and $\textbf{s}_{||}$ defines energy transfer in a longitudinal wave. This reflects the fact that the given system of equations describes simultaneously propagation of both longitudinal and transverse waves. Let us note here that the density of the total energy $\epsilon$ is positive definite quantity.  

Let us consider now the question of interaction of a triple of fields $\textbf{E}$, $\textbf{H}$ and $W_{FP}$ with charges. Since all the three fields are expressed via 4-potential components according to formulas (\ref{ap3}) and (\ref{ap4}), we can chose a standard Lagrangian \cite{LLTF} as a Lagrangian describing this interaction: 
\begin{eqnarray}
L=\frac{m\textbf{v}^2}{2}+\frac{1}{c}(q\textbf{A}\textbf{v})-q\varphi,
\label{ap11}
\end{eqnarray}
where $q$, $m$ and $\textbf{v}$ - are the charge, mass and velocity of the charged body, correspondingly. The equations of motion following from (\ref{ap11}) have the form:
\begin{eqnarray}
m\frac{d\textbf{v}}{dt}=q\textbf{E}+\frac{q}{c}\left[\textbf{v}\times\textbf{H}\right].
\end{eqnarray}
Here the field $W_{FP}$ is not included into the right side. The law of variation of kinetic energy of a charged body, following from these equations, is defined by the expression: 
\begin{eqnarray}
\frac{d}{dt}\frac{m\textbf{v}^2}{2}=\textbf{IE},
\label{ap13}
\end{eqnarray}
where $\textbf{I}$ is the total current: $\textbf{I}=q\textbf{v}$. For definition of the total energy conservation law  it is necessary in the formula (\ref {ap9}) to pass from charge density to a point charge and integrate. 
As a result we get:
\begin{eqnarray}
\frac{d}{dt}\Xi=-\textbf{IE}+cq\,W_{FP},
\label{ap14}
\end{eqnarray}
where $\Xi$ is the total energy of the fields. Summing up (\ref{ap13}) and (\ref{ap14}), the following can be obtained:
\begin{eqnarray}
\frac{d}{dt}\left[\Xi+\frac{d}{dt}\frac{m\textbf{v}^2}{2}\right]=cq\,W_{FP}.
\label{ap15}
\end{eqnarray}
From this formula it is seen that the total energy of a charged body interacting with the fields $\textbf{E}$, $\textbf{H}$ and $W_{FP}$ is not kept.

Another interesting peculiarity of this theory is the behavior of the fields at gauge transformations of the potentials \cite{Dzhekson}:
\begin{eqnarray}
\varphi'=\varphi-\frac{1}{c}\frac{\partial\,f}{\partial\,t},\quad\quad \textbf{A}'=\textbf{A}+\nabla\,f,
\label{ap16}
\end{eqnarray}
where $f$- is an arbitrary gauge function. In this case the fields $\textbf{E}$ and $\textbf{H}$ are invariant relative to these transformations, while the field $W_{FP}$ transforms according to the law:
\begin{eqnarray}
W_{FP}'=\frac{1}{c}\frac{\partial\varphi'}{\partial\,t}+{\rm div}\textbf{A}'=W_{FP}-\frac{1}{c^2}\frac{\partial^2f}{\partial\,t^2}+\Delta\,f.
\label{ap17}
\end{eqnarray}
However, Lagrangian (\ref{ap1}) can be made gauge-invariant if it is demanded that along with  transformations for potentials (\ref{ap16}) the densities of charges and currents be transformed according to the law:
\begin{eqnarray}
\rho'=\rho+\frac{1}{4\pi\,c}\frac{\partial}{\partial\,t}\left(-\frac{1}{c^2}\frac{\partial^2f}{\partial\,t^2}+\Delta\,f\right),\quad\quad            \textbf{j}'=\textbf{j}-\frac{c}{4\pi}\nabla\left(-\frac{1}{c^2}\frac{\partial^2f}{\partial\,t^2}+\Delta\,f\right).
\end{eqnarray}
However, it is necessary to note that the law of charge conservation places a limitation on the gauge function  $f$. To all intents, the continuity equation must be gauge-invariant and, as a consequence, $f$ must satisfy the equation:
\begin{eqnarray}
-\frac{1}{c^2}\frac{\partial^2}{\partial\,t^2}\left(-\frac{1}{c^2}\frac{\partial^2f}{\partial\,t^2}+\Delta\,f\right)+\Delta\left(-\frac{1}{c^2}\frac{\partial^2f}{\partial\,t^2}+\Delta\,f\right)=0.
\end{eqnarray}
Let us note in conclusion that the considered theory of electromagnetic field on the basis of the Fock-Podolsky Lagrangian describes propagation of both transverse electromagnetic and longitudinal waves in vacuum in the nonrelativistic case. However, this theory contains an obstacle associated with the law of total energy conservation (\ref{ap15}), and this obstacle can be eliminated only if it is demanded that $W_{FP}=0$ (in this case we pass to the classical Maxwell theory in which there is no longitudinal wave in vacuum). One more weakness of this theory lies, in our opinion, in the transformational properties of the system of fields and, to all intents, by its definition (\ref{ap4}) the field $W_{FP}$ is Lorentz-invariant, i.e. it does not change at Lorentz transformations while the fields $\textbf{E}$ and $\textbf{H}$ transform via each other. At gauge transformations, by contrast, it is the fields $\textbf{E}$ and $\textbf{H}$, that are invariant and $W_{FP}$ transforms according to law (\ref{ap17}). In other words,  despite the fact that Lagrangian (\ref{ap1}) is Lorentz-invariant and can satisfy the condition of gauge invariance, the fields $\textbf{E}$, $\textbf{H}$ and $W_{FP}$ do not form a single physical object  and, therefore, introduction of longitudinal waves in Maxwell's electrodynamics on the basis of the Fock and Podolsky Lagrangian is on the whole unsatisfactory. 

\end{document}